\documentstyle[sprocl]{article}

\input{epsf}

\def\Journal#1#2#3#4{{#1} {\bf #2}, #3 (#4)}

\def\NPB{{\em Nucl.\ Phys.}\ B}
\def\PLB{{\em Phys.\ Lett.}\  B}
\def\PRL{\em Phys.\ Rev.\ Lett.\ }
\def\PRD{{\em Phys.\ Rev.}\ D}
\def\ZPC{{\em Z. Phys.}\ C}
\def\EPJC{{\em Eur.\ Phys.\ J.} C}
\def\JPG{{\em J. Phys.}\ G}

\begin{document}

\title{\vspace{-2em} {\hfill\normalsize hep-ph/9909445} \\ \vspace{2em}
POLARISATION IN \\ DEEPLY VIRTUAL MESON PRODUCTION\footnote{Talk given at
    the Workshop on Exclusive and Semiexclusive Processes at High
    Momentum Transfer, Jefferson Lab, Newport News, VA, USA, 20--22
    May 1999, to appear in the proceedings.}}

\author{M. DIEHL} \address{Deutsches Elektronen-Synchroton DESY, 22603
  Hamburg, Germany\\[0.5ex]
  present address: \\
  Stanford Linear Accelerator Center, P.O.~Box 4349, 
  Stanford, CA 94309, U.S.A.}

\author{T. GOUSSET} \address{SUBATECH, B.P. 20722, 44307 Nantes,
  France} 

\author{B. PIRE} \address{Centre de Physique Th\'eorique, \'Ecole
  Polytechnique, 91128 Palaiseau, France}

\maketitle\abstracts{We discuss two aspects of polarisation in hard
  exclusive meson production: the leading-twist selection rule for the
  meson helicity, and the different partial waves of a $\pi\pi$-pair
  which may or may not be due to the decay of a $\rho$.}

\section{Helicity in vector meson production}

\begin{figure}[t]
\begin{center}
  \leavevmode
  \epsfxsize 0.6\textwidth
  \epsfbox{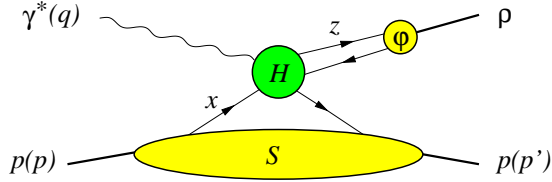}
\end{center}
\caption{{}\label{fig:factorise} Factorisation of exclusive
  $\rho$-production into a hard-scattering coefficient $H$, a skewed
  quark distribution $S$ and a meson distribution amplitude $\varphi$.
  There are also diagrams with the skewed gluon instead of the quark
  distribution. $x$ and $z$ denote momentum fractions in the proton
  and the $\rho$, respectively.}
\end{figure}

The first half of this talk is about exclusive electroproduction of a
$\rho$ (or any other light vector meson), $ep \to ep + \rho$, in the
Bjorken limit where the photon virtuality $Q^2 = -q^2$ becomes large,
while $x_B = Q^2 /(2 p\cdot q)$ and the invariant momentum transfer $t
= (p-p')^2$ remain fixed. There is a factorisation
theorem\,\cite{factoriz} stating that in this limit the $\gamma^* p$
amplitude factorises into a hard photon-parton scattering and
non-perturbative quantities, namely skewed quark and gluon
distributions in the proton and a $q\bar{q}$ distribution amplitude of
the meson. This is shown in Fig.~\ref{fig:factorise}, where also
four-momenta are defined.

The factorisation theorem provides a solid basis for extracting
information on the quark and gluon structure of hadrons from this
process, in particular for measuring skewed parton distributions. The
description of the amplitude in terms of these quantities is of course
only accurate up to power corrections in $1/Q$, and in a data analysis
it is essential to investigate how close one is to the asymptotic
regime at a given value of $Q^2$. It is at this point that helicity
selection rules are of great value.

An essential part of the factorisation theorem is that the description
of Fig.~\ref{fig:factorise} only holds if the photon is longitudinally
polarised in the $\gamma^* p$ c.m. The corresponding amplitude has a
scaling behaviour ${\mathcal A}(\gamma^*_L p) \sim {\mathrm const}/Q$
in the Bjorken
limit. For a transverse photon factorisation cannot be established
since the loop integrals in the relevant Feynman graphs are sensitive
to dangerous infrared regions, but the theorem predicts the
corresponding amplitude to be suppressed by at least one power,
${\mathcal A}(\gamma^*_T p) \sim {\mathrm const}/Q^2$.

It turns out that there is a second selection rule, stating that to
leading power in $1/Q$ one can only produce a longitudinal vector
meson, while transverse $\rho$ production is again power suppressed.
Collins, Frankfurt and Strikman\,\cite{factoriz} observed that a
transition $\gamma^*_L p \to \rho_T^{\phantom{*}} p$ at order $1/Q$
must involve the so-called chiral-odd quark distribution in the
proton and the chiral-odd distribution amplitude of the $\rho$. This
was promising given that experimental information on these quantities
is very scarce. Unfortunately, the corresponding hard-scattering
coefficient $H$ was found to be zero to lowest order in the strong
coupling.\cite{mpw} A general symmetry argument shows that it is in
fact zero to all orders in $\alpha_s$.\cite{dgp}

Let us see what goes into this argument. The first ingredient is that
the leading power behaviour of the diagrams in
Fig.~\ref{fig:factorise} is obtained by replacing the relative
transverse momenta of the partons with zero when evaluating the hard
scattering $H$. The hard subprocess is therefore collinear, say along
the $z$-axis. The loop integral over the transverse momenta is
performed in the soft quantities $S$ and $\varphi$ alone, and from
rotation invariance it is easy to see that in $\varphi$ the helicities
of the quark and antiquark must add up to the helicity of the meson.
For a transverse $\rho$ this gives a chiral-odd distribution
amplitude, i.e., one where $q$ and $\bar{q}$ have opposite chirality
and thus equal helicity. The second ingredient is that when
calculating to leading-power accuracy one can set the quark mass to
zero in $H$. Then the hard scattering conserves quark helicity and
thus selects the chiral-odd quark distribution in the proton target;
chiral-even quark distributions and the gluon distribution do not
contribute. When one follows the flow of helicity from the incoming to
the outgoing quark lines in the Feynman graphs for $H$ one finds that
the angular momentum along the $z$-axis changes by two units, a
mismatch that cannot be compensated by the photon polarisation. Thus
one finds $H=0$.

The previous argument did not take into account that at higher orders
in $\alpha_s$ the Feynman diagrams for $H$ have infinities, so that
one must first regularise them, then perform appropriate subtractions
of these divergences, and finally remove the regularisation. Now, as
soon as one regularises the theory, chirality is no longer conserved.
Chirality breaking terms can survive even after the regularisation has
been removed again, which gives for instance rise to the axial anomaly
in QCD. Hoodbhoy and Lu\,\cite{hl} have calculated the one-loop
diagrams for $H$ and found indeed a non-zero result.

Their result is however incomplete since it misses the subtractions of
collinear divergences that have to be made in the calculation of a
hard-scattering coefficient $H$; if these are included one will find
$H=0$ again. One can in fact give a general proof\,\,\cite{cd} that to
any finite order in perturbation theory the hard-scattering
coefficient conserves the chirality of its external quark lines, so
that the symmetry argument sketched above is valid. The idea of the
proof is to regularise the theory by going to $4-\epsilon$ dimensions,
perform all subtractions of divergences and then set $\epsilon=0$,
leaving the helicities of the external quarks in $H$ unspecified. The
final result for $H$ lives in 4 dimensions, where chirality
conservation it ensured by having an odd number of Dirac matrices
$\gamma^\mu$ along each quark line. The latter is a consequence of the
Feynman rules of massless QCD and is not invalidated by any of the
intermediate steps in $4-\epsilon$ dimensions.

An alternative would be to use Pauli-Villars regularisation to render
the loop integrals finite. Then chirality remains conserved along
massless fermion lines. Only in internal quark loops do massive
regulator fermions occur whose chirality is not conserved; this leads
for instance to the axial anomaly in the triangle diagram of two
vector and one axial current. It does, however, not disturb the fact
that chirality is conserved for the external quark lines of $H$, which
remain massless.

It is important to note that the argumentation leading to our helicity
selection rule is very similar to the one that establishes hadron
helicity conservation for exclusive processes at leading
twist.\cite{bl} The proof that a hard-scattering coefficient in
perturbative QCD conserves chirality carries over to this case.

In summary, we have that the only leading-twist helicity amplitude is
the one for $\gamma^*_L p \to \rho_L^{\phantom{*}} p$. All others must
be due to power corrections. As the angular distributions in $ep \to
ep + \rho \to ep + \pi^+ \pi^-$ contain detailed information on the
various helicity transitions, their study can give valuable
information on how far one is from the asymptotic regime where the
helicity selection rules apply.

Beyond this they may guide theory in describing the physics of power
corrections itself, a field where we are far yet from a systematic
theory. A crucial ingredient in deriving our selection rule was the
collinearity of the hard scattering subprocess, and as soon as one
keeps transverse parton momenta (and hence $t$) finite there, all
helicity combinations appear. The relevance of the transverse momentum
distribution of partons within hadrons has been emphasised in several
papers dealing with the helicity structure of meson production in the
small-$x_B$ regime,\cite{mrt-ik,nik,cud-roy} where increasingly
accurate data are becoming available.\cite{h1-zeus} Another important
question is
that of perturbative control over infrared regions in the loop
integrals, to which the different helicity amplitudes are sensitive to
different degrees (remember what we said about $\gamma^*_L$ and
$\gamma^*_T$ in the beginning).

\section{From $\rho$ to $\pi\pi$}

So far we have looked at $\gamma^* p \to \rho\, p$ and its description
by the factorised diagrams of Fig.~\ref{fig:factorise}, having in the
back of our minds that in practice one observes the $\rho$ via its
decay channel $\rho\to \pi^+\pi^-$. In the second part of this talk we
will take a different point of view on the same process and directly
describe $\gamma^* p \to \pi^+\pi^- + p$ in a factorised way.  That
is, we replace in Fig.~\ref{fig:factorise} the distribution amplitude
$\varphi(z)$ of the $\rho$ by a generalised distribution amplitude
(GDA), which describes the non-perturbative transition from a
$q\bar{q}$-pair, produced in the hard scattering, to the final state
$\pi^+\pi^-$.\cite{GDA-old,dgpt} The proof of factorisation carries
over including its fine print,\cite{af} and the amplitude for
$\gamma^* p \to \pi^+\pi^- + p$ has the same scaling behaviour and
photon helicity selection rule as the one for $\gamma^* p \to \rho\,
p$. To formulate the second selection rule we replace the helicity of
the $\rho$ by the projection $l_z$ of the $\pi\pi$ angular momentum
along the $z$-axis (defined as opposite to the momentum $p'$ in the
$\pi\pi$ c.m.): To leading power in $1/Q$ one has $l_z=0$.

It is very natural that such a generalised factorisation holds: the
transition from $q\bar{q}$ to the pion pair is all long-range physics,
and the formation and decay of a $\rho$ resonance are just different
parts of this. Maybe even more important is that the transition also
receives contributions from the nonresonant $\pi\pi$ continuum. In the
context of factorisation between short and long distances (and thus
for instance in the study of skewed parton distributions) it is not
necessary to separate a resonance ``signal'' from a continuum
``background''. In fact one need not even require the invariant mass
$M$ of the pion pair to be close to the $\rho$ mass; all that counts
for factorisation is that $M^2$ be small compared to the large scale
$Q^2$.

Factorisation not only extends to invariant masses off the $\rho$
peak, but also to $\pi\pi$ partial waves other than the $P$-wave
corresponding to $\rho$-decay (each in its $l_z=0$ projection
according to our selection rule). To discuss this in more detail we
list the kinematical variables on which a GDA $\Phi$ depends: There is
the invariant mass $M$, the quark momentum fraction $z$ with respect
to the total momentum of the pair (in a frame where the pair moves
fast), and the polar angle $\theta$ of the $\pi^+$ in the $\pi\pi$
c.m.\ (with the $z$-axis defined as above). Expanding
$\Phi(z,\cos\theta,M^2)$ in Legendre polynomials $P_l(\cos\theta)$ one
obtains the partial wave expansion of the pion system. Odd partial
waves ($l=1,3,\ldots$) have negative charge conjugation parity $C$,
just as mesons of the $\rho$ family, while the even ones
($l=0,2,4,\ldots$) have positive $C$-parity and thus the quantum
numbers of $f$-mesons. It is useful to project out the $C$-even and
odd parts of $\Phi$ according to $\Phi^{\pm}(z,\cos\theta,M^2) =
\frac{1}{2} [\, \Phi(z,\cos\theta,M^2) \pm \Phi(z,-\cos\theta,M^2)\,
]$.

Just as ordinary distribution amplitudes, $\Phi$ also depends on a
factorisation scale $\mu$, and this dependence is described by the
usual ERBL evolution equations.\cite{ERBL} For pion pairs in the
$\rho$-channel they have the asymptotic solution
\begin{equation}
\Phi^- \stackrel{\mu\to\infty}{\sim} {\mathrm const}\
  z(1-z) \, F_{\pi}(M^2) \, \beta \cos\theta ,
\label{asy}
\end{equation} 
where $\beta=\sqrt{1-4 m_\pi^2 /M^2}$ is the velocity of the pions in
the $\pi\pi$ c.m.\ and $F_{\pi}(M^2)$ the timelike pion form factor,
measured in $e^+e^-\to \pi^+\pi^-$. Note that even in the completely
asymptotic regime the mass distribution is governed by the pion form
factor, which contains the $\rho$ resonance \emph{and} the $\pi\pi$
continuum. While there are certainly non-factorising contributions to
the amplitude that lead to a distortion of the mass spectrum away from
a resonance shape, the converse is not true: factorisation does
\emph{not} predict the $\pi\pi$ mass spectrum to become a pure
Breit-Wigner form for the $\rho$.

While the asymptotic form (\ref{asy}) of $\Phi^-$ is a pure $P$-wave,
the asymptotic form of $\Phi^+$ has a $z$-dependence $z(1-z)(2z-1)$
and contains an $S$- and a $D$-wave, each multiplied with an
$M^2$-dependent form factor. Higher partial waves, $l\ge 3$, occur
only to the extent that the GDAs deviate from their asymptotic
forms.\cite{Poly}

In the process $\gamma^* p \to \pi^+\pi^- + p$ both $C$-even and odd
pion pairs can be produced, so that in general one has a coherent
superposition of pairs with different quantum numbers, even when $M$
is close to or on the $\rho$ mass peak. Observable consequences of the
presence of $f$-type pairs are that
\begin{list}{$\bullet$}{\renewcommand{\leftmargin}{1.2em}
    \renewcommand{\itemsep}{0em} \renewcommand{\parsep}{0em}}
\item the angular distribution of $\pi^+\pi^-$ is not the one of a
  pure $P$-wave. For analyses of the helicity density matrix in
  $\rho$-production it is essential to know how important
  contributions from other partial waves, e.g.\ the $S$- and $D$-waves
  are, since the usual extraction methods for the $\rho$ helicity
  density matrix assume a pure $P$-wave.
  
  It is worth noting that the interference between even and odd
  partial waves gives a contribution to the $\pi^+\pi^-$ angular
  distribution that is odd under the exchange of the four-momenta of
  $\pi^+$ and $\pi^-$. It should be easy to test whether such $C$-odd
  terms are present in the angular distribution, e.g.\ by taking
  moments of $C$-odd functions like $\cos\theta$ or $\sin\theta
  \cos\varphi$.
\item not only $\pi^+\pi^-$ but also $\pi^0\pi^0$-pairs are produced.
  In fact, isospin invariance tells us that the GDA for $\pi^0\pi^0$
  is equal and opposite to $\Phi^+$ for $\pi^+\pi^-$.
\end{list}  

The production of $\pi\pi$-pairs with different $C$-parity involves
exchanges with different quantum numbers in the $t$-channel, namely
$C$-plus exchange for $C$-odd pairs and $C$-minus exchange for
$C$-even pairs.
In the Bjorken limit this means different combinations of skewed
parton distributions. For $\rho$-type pairs one is sensitive to the
quark and the gluon distributions, while for $f$-type pairs quarks
contribute but gluons do not: $C$-minus exchange requires at least three
gluons in the $t$-channel, which gives an amplitude that is power
suppressed relative to the leading $1/Q$ of quark-antiquark or
two-gluon exchange.\cite{factoriz} It is beyond the scope of our
investigation to estimate how important this nonleading-twist
contribution becomes when $Q^2$ is not so large.

To leading-twist accuracy and leading order in $\alpha_s$ the ratio of
the amplitudes for $f$- and $\rho$-channel pion pairs is
\begin{eqnarray}
\lefteqn{\frac{\mathcal{A}(\mbox{$C$-even pairs})}{
               \mathcal{A}(\mbox{$C$-odd pairs})} =
  \frac{\int_0^1 dz\, \frac{2z-1}{z(1-z)}\, \Phi_u^+}
       {\int_0^1 dz\, \frac{1}{z(1-z)}\, \Phi_u^-}}   \nonumber \\
&& \times
  \frac{\int_0^1 dx\, \frac{1}{(x-\xi+i\epsilon)(x+\xi)}\,
        \Big( \frac{2}{3} \xi (H_u - H_{\bar{u}}) -
              \frac{1}{3} \xi (H_{d\phantom{\bar{d}}\!\!\!} - 
                               H_{\bar{d}}) \Big) + \ldots}
       {\int_0^1 dx\, \frac{1}{(x-\xi+i\epsilon)(x+\xi)}\,
        \Big( \frac{2}{3} x (H_u + H_{\bar{u}}) +
              \frac{1}{3} x (H_{d\phantom{\bar{d}}\!\!\!} + 
                             H_{\bar{d}}) +
              \frac{3}{4} H_g \Big) + \ldots}
  \label{ratio} 
\end{eqnarray}
in the case where the proton helicity is not flipped.\footnote{The
amplitude for $f$-channel pair production contains a further term
going with the GDA for the transition from two gluons to $\pi\pi$, as
remarked by Lehmann-Dronke {\it et al.}\,\protect\cite{LEH} With their
ansatz for the GDAs they estimate that this contribution, missing in
(\protect\ref{ratio}), may be twice as large as the one with
$\Phi_u^+$.} The dots stand
for terms with the skewed distributions $E_q$, $E_{\bar{q}}$, $E_g$;
they are the only ones that contribute in the case of proton helicity
flip. We use skewed quark distributions $H_q(x,\xi,t)$ as defined by
Ji,\cite{ji} antiquark distributions $H_{\bar{q}}(x,\xi,t)
=-H_q(-x,\xi,t)$, and $H_g(x,\xi,t) = 2x H_g^{\mathrm{Ji}}(x,\xi,t)$
for the gluons. The non-skewed limits of these functions are
$H_q(x,0,0)=q(x)$, $H_{\bar{q}}(x,0,0)=\bar{q}(x)$ and $H_g(x,0,0)=x
g(x)$. In (\ref{ratio}) we have neglected the formation of a pion pair
from $s\bar{s}$, and used isospin invariance to relate the GDAs for
$d$- and $u$-quarks, $\Phi^{\pm}_d = \pm\,
\Phi^{\pm}_{u\phantom{d}\!\!\!}$.

A number of theory predictions for skewed parton distributions can be
found in the literature, and one may use one's favourite model in
order to evaluate the ratio of integrals over skewed parton
distributions. For a very crude order-of-magnitude estimate we replace
the second line of Eq.~(\ref{ratio}) with the ratio
\begin{equation}
\frac{\frac{2}{3} (u-\bar{u}) - \frac{1}{3} (d-\bar{d})}
     {\frac{2}{3} (u+\bar{u}) + \frac{1}{3} (d+\bar{d}) +
      \frac{3}{4} g}
  \label{approx}
\end{equation}
of the usual parton densities, evaluated at a momentum fraction of
order $\xi = x_B /(2-x_B)$. As a numerical example we take the GRV~LO
parametrisation\,\cite{grv} at a factorisation scale
$\mu^2=4~\mbox{GeV}^2$ and find that the ratio (\ref{approx}) changes
from 0.15 to 0.5 for momentum fractions from 0.1 to 0.4. We thus
expect the second line of Eq.~(\ref{ratio}) to be small at values of
$x_B$ where gluons dominate over quarks, whereas for $x_B$ in the
valence region it may be of order~1.

To estimate the ratio of integrals over GDAs in Eq.~(\ref{ratio}) we
take the asymptotic solution (\ref{asy}) in the $\rho$-channel, which
is completely determined given our knowledge of the pion form factor.
The asymptotic GDA in the $f$-channel involves two form factors which
are unknown at the values of $M$ where we need them. Indirect
information on them can be obtained from crossing symmetry, which
relates GDAs to the parton distributions in the pion. Using the
results of Polyakov\,\cite{Poly} and making a number of
approximations\,\cite{dgp-prep} we obtain for $M$ below 1~GeV
\begin{equation}
\frac{\int_0^1 dz \frac{2z-1}{z(1-z)}\, \Phi_u^+}
       {\int_0^1 dz \frac{1}{z(1-z)}\, \Phi_u^-} \approx
  - \frac{ \frac{5}{3} R_q [\, (3-\beta^2) \exp(i\delta_S) -
           \beta^2\, (3\cos^2\theta - 1) \exp(i\delta_D) \,] }
         { 6\beta\, \cos\theta \exp(i\delta_P)\; |F_\pi(M^2)| } ,
  \label{GDA-frac}
\end{equation}
where $R_q$ denotes the fraction of the pion momentum carried by
quarks and is between $0.6$ and $0.5$ for the GRS~LO parton
distributions in the pion\,\cite{grs} at a factorisation scale $\mu^2$
between 1 and 20 GeV$^2$.  $\delta_S$, $\delta_P$ and $\delta_D$ are
the phase shifts for $\pi\pi$ elastic scattering in the appropriate
partial waves, which are rather well known for $\pi\pi$ invariant
masses below 1 GeV. A determining factor in (\ref{GDA-frac}) is
$|F_\pi(M^2)|$, which has a value around 1.4 at $M = 400$~MeV and
around 1.8 at $M = 1000$~MeV, but is as large as 6.7 at the $\rho$
mass peak.

In conclusion, we estimate that the relative contribution from
$f$-channel pion pairs in electroproduction should be rather small for
invariant masses $M$ around the $\rho$-peak. A few 100 MeV off peak,
however, it may well be of importance in the $x_B$-region where gluon
exchange does not dominate over quarks, and especially the
interference between $f$- and $\rho$-channel pairs may be visible. We
also note that in (\ref{GDA-frac}) not only $|F_\pi|$ has a strong
dependence on $M^2$ but also the $\pi\pi$ phase shifts, so that the
different contributions to the pion angular distribution will have a
marked $M^2$-behaviour.

We finally remark that there are other pion pair production processes
where the $\pi\pi$ formation is described by GDAs. One example is
$\gamma^*\gamma \to \pi\pi$ with the photon virtuality much larger
than the $\pi\pi$ invariant mass; in this case $\Phi^+$ but not
$\Phi^-$ contributes.\cite{dgpt} Another is photoproduction, $\gamma p
\to \pi^+\pi^- + Y$, where the proton dissociates into a hadronic
system $Y$ with invariant mass much smaller than the $\gamma p$ c.m.\ 
energy, and where the momentum transfer $t$ between $p$ and $Y$ is
large. Data for this reaction are beginning to come in.\cite{rho-exp}
At high energies it is dominated by two-gluon exchange in the case
where the pion pair is due to the decay of a $\rho$. A fair amount of
theory has been worked out for this.\cite{rho-thy} It is important to
realise that the production of $f$-type pion pairs through three-gluon
exchange has the \emph{same} scaling behaviour in $t$ as the
production of $\rho$-type pairs, i.e., it is not power suppressed as
in electroproduction at small $t$.\cite{semi} Away from the mass peak
of the $\rho$ the contribution from $f$-channel pairs may therefore be
important.

\section*{Acknowledgements}

This work has been partially funded by the TMR programme of the
European Union, Contracts No.~FMRX-CT96-0008 and No.~FMRX-CT98-0194.
SUBATECH is Unit\'e mixte 6457 de l'Universit\'e de Nantes, de l'Ecole
des Mines de Nantes et de l'IN2P3/CNRS. Centre de Physique Th\'eorique
is Unit\'e mixte C7644 du CNRS.

\end{document}